\documentclass[fleqn,12pt]{article}
\usepackage{amsmath}
\usepackage{amsfonts,latexsym,stmaryrd}
\numberwithin{equation}{section}
\newcommand{\ii}{\mathrm{i}}

\newcommand{\pd}{\partial}
\newcommand{\dd}{\mathrm{d}}

\newcommand{\e}{\mathrm{e}}
\newcommand{\ket}[1]{\left|#1\right\rangle}
\newcommand{\bra}[1]{\left\langle #1\right|}

\newcommand{\I}{\mathbb{I}}

\newcommand{\ft}[2]{{\textstyle\frac{#1}{#2}}}
\def\tilde{\widetilde}
\def\1bar{1\hskip -.275cm -}
\def\2bar{2\hskip -.275cm -}
\def\3bar{3\hskip -.275cm -}

\newsavebox{\uuunit}

\newcommand{\nn}{\nonumber}

\newcommand{\nc}{\newcommand}
\nc{\la}{\lambda} \nc{\alf}{\alpha} \nc{\tht}{\theta}
\nc{\eps}{\epsilon} \nc{\ga}{\gamma} \nc{\Ga}{\Gamma}
\nc{\De}{\Delta} \nc{\de}{\delta} \nc{\si}{\sigma}
\nc{\ka}{\kappa} \nc{\om}{\omega} \nc{\qq}{\quad\quad}
\nc{\nf}{\infty} \nc{\dl}{\mathop{\smash{\cal L}}}
\nc{\ol}{\overline} \nc{\beq}{\begin{equation}}
\nc{\barr}{\begin{array}} \nc{\earr}{\end{array}}
\nc{\eeq}{\end{equation}} \nc{\beqa}{\begin{eqnarray}}
\nc{\dst}{\displaystyle}\nc{\pt}{\partial}
\nc{\eeqa}{\end{eqnarray}} \nc{\nnb}{\nonumber}
\nc{\bs}{\backslash}        \nc{\mbb}{\mathbb}
\nc{\brm}{\begin{remunerate}} \nc{\erm}{\end{remunerate}}
\nc{\vareps}{\varepsilon} \nc{\tb}{\tilde\beta_0} \nc{\ts}{\tilde
s} \nc{\tth}{\tilde \theta}
\newcounter{muni}
\usecounter{muni}
  \nc{\lapdec}{\mathop{\Delta}}

\def\theequation{\arabic{section}.\arabic{equation}}
\newenvironment{remunerate}{\begin{list}{{\rm \arabic{muni}.}}
{\usecounter{muni}
\setlength{\leftmargin}{0pt}\setlength{\itemindent}{38pt}}}{\end{list}}
\newcommand{\alg}[1]{\mathfrak{#1}}

\newcommand{\alG}{\alg{g}}

\newcommand{\alSU}{\alg{su}}

\newcommand{\alPSU}{\alg{psu}}

\nc{\cre}{\color[rgb]{1.00,0.00,0.00}}
\nc{\cgr}{\color[rgb]{0.00,1.00,0.00}}

\newcommand{\dg}{^\dagger}

\newcommand{\be}{\begin{equation}} \newcommand{\ee}{\end{equation}}
\newcommand{\bea}{\begin{eqnarray}} \newcommand{\eea}{\end{eqnarray}}
\newcommand{\ben}{\begin{displaymath}}
\newcommand{\een}{\end{displaymath}}

\makeatletter
\def\hlinewd#1{%
\noalign{\ifnum0=`}\fi\hrule \@height #1 %
\futurelet\reserved@a\@xhline} \makeatother

\begin{document}

\title{Superstring sigma models from spin chains:\\
the SU$(1,1|1)$ case}

\author{ {\bf S. Bellucci, P.-Y. Casteill} \\
{\it INFN -- Laboratori Nazionali di Frascati,}\\
{\it Via E. Fermi 40, 00044 Frascati, Italy} \\
 {\bf J.F. Morales} \\
{\it Department of Physics, CERN  Theory Division}\\
{\it 1211 Geneva 23, Switzerland}
}

\maketitle
\begin{abstract}
We derive the coherent state representation of the integrable spin
chain Hamiltonian with non-compact supersymmetry group
$G=$SU$(1,1|1)$. By passing to the continuous limit, we find a
spin chain sigma model describing a string moving on the
supercoset $G/H$, $H$ being the stabilizer group. The action is
written in a manifestly $G$-invariant form in terms of the Cartan
forms and the string coordinates in the supercoset.

 The spin chain sigma model is shown to agree with that
following from the Green-Schwarz action describing two-charged string spinning
on  AdS$_5\times S^5$.

\end{abstract}
\newpage

\tableofcontents
% ----------------------------------------------------------------
\section{Introduction}

% ----------------------------------------------------------------

The AdS/CFT correspondence
\cite{Maldacena:1998re,Gubser:1998bc,Witten:1998qj} between
strings on anti-de Sitter (AdS) spaces and boundary gauge theories is now of common
use.
 The typical example relates string theory on AdS$_5\times S^5$ to
 ${\cal N}=4$ supersymmetric Yang-Mills (SYM).
 String states in the bulk are in correspondence to gauge invariant operators
 in the boundary and likewise correlation functions in the two theories
 are related by a well established holographic dictionary.
  There are several tests of the correspondence at the supergravity level
  (see \cite{Aharony:1999ti} for a review and a complete list of
references), but few ones beyond this limit. Waiting for a better
understanding of the string physics on AdS one can explore
particular limits of the AdS geometry where physics simplifies itself. In
\cite{Bianchi:2003wx,Beisert:2003te,Beisert:2004di,Morales:2004xc}
the spectrum of strings on AdS and SYM operators was shown to
agree at the higher symmetry enhancement point. In
\cite{Berenstein:2002jq}, the authors explore the holographic
correspondence in the neighborhood of null geodesics of
AdS$_5\times S^5$, where the geometry looks like a pp-wave
\cite{Blau:2001ne}. String theory on pp-wave geometries is known
to be solvable \cite{Metsaev:2001bj,Metsaev:2002re}.
 On the gauge theory side this limit corresponds to focusing on SYM operators
 with large ${\cal R}$-symmetry charge $J$.

 In a similar spirit, fluctuations around semiclassical spinning
 strings were studied in \cite{Gubser:2002tv}\nocite{Frolov:2002av,
Tseytlin:2003ac,Frolov:2003xy,Frolov:2003qc,Frolov:2003tu,Frolov:2004bh,
Arutyunov:2003uj,Beisert:2003ea}-\cite{Beisert:2003xu}. Once again,
energies of classical string solutions were shown to match the
anomalous dimensions of SYM operators with large charges.
On the gauge theory side, the planar one-loop anomalous dimensions
in ${\cal N}=4$ SYM are governed by integrable spin
chain Hamiltonians \cite{Minahan:2002ve,Beisert:2003yb,
Beisert:2003jj}.
 Non planar corrections were computed in
 \cite{Beisert:2002ff}\nocite{Bellucci:2004ru}
 -\cite{Bellucci:2004qx} in terms of a joining-splitting spin
 chain operator mimicking string
interactions. An alternative approach to the description of non
planar corrections can be found in \cite{Bellucci:2004cs}.
Moreover, the analysis of \cite{Bellucci:2004ru}
 -\cite{Bellucci:2004qx} was extended in \cite{Bellucci:2004am} to the two loop level of SYM
 perturbation theory, by considering the SYM anomalous dimension/mixing matrix to two loops
and applying to it the map to the spin bit system.
In the large $N$ limit, the corresponding SU(2) spin bit model was shown to be reduced to the two loop planar integrable spin chain
\cite{Bellucci:2004am}.

In the continuous (BMN) limit, i.e. for SYM operators with large
$J$, the spin chain can be identified with the worldsheet of a
closed string with spin chain excitations describing the string
profile in the symmetry group taken as a target space. The spin chain
Hamiltonian describes the dynamics of this string. As for the BMN
case, the perturbative regime of SYM is accessible to this limit
and accordingly the string and spin chain sigma model actions
should agree.
 This was shown to be the case in
 \cite{Kruczenski:2003gt}
\nocite{Kruczenski:2004cn,Hernandez:2004uw,Kristjansen:2004za,
Kruczenski:2004cn}-\cite{Kruczenski:2004kw}
 for $SO(6)$ and its compact subgroups, in
 \cite{Stefanski:2004cw,Bellucci:2004qr,Ryang:2004pu} for SL$(2)$ and recently
  in \cite{Hernandez:2004kr} for the compact supergroup SU$(1|2)$.
  In all cases, semiclassical spinning string states were identified
with coherent states made out of spin chain eigenstates of the
symmetry group (see \cite{Kazakov:2004nh}
\nocite{Kruczenski:2004wg,Arutyunov:2004yx,Park:2005ji,Khan:2005fc,Berenstein:2005fa,Freyhult:2005fn,Beisert:2005mq}
-\cite{Hernandez:2005nf} for further developments in this subject).

 The aim of this note is to extend this result to the simplest non-compact supergroup, namely
 $G=$SU$(1,1|1)$. This sector corresponds to SYM operators made out of a
single scalar, a fermion and its derivatives along a fixed
direction. It gives the minimal supersymmetric extension of the
SL$(2)$ spin chain. On the string side they describe
supersymmetric excitations around a string spinning in both $S^5$
and AdS$_5$. We first derive the coherent state representation of
the spin chain Hamiltonian. SU$(1,1|1)$ is non-compact and
its representations are infinite dimensional.
 This makes the analysis of SU$(1,1|1)$ more involved
 than for the $SU(1|2)$ case and leads to a non-linear form for
 the Hamiltonian.
 Remarkably, like in the SL$(2)$ bosonic case
 \cite{Stefanski:2004cw,Bellucci:2004qr},
 the infinite series of "higher derivative" terms
 can be summed into a simple Log dependence. By passing to the continuous limit, the spin chain
  action  reduces to a linear sigma model for a string moving on the
  supercoset SU$(1,1|1)/$SU$(1|1)\times $U$(1)$. The results in this limit
  (and only in this limit) are
  related to those for SU$(1|2)$ found in \cite{Hernandez:2004kr} via an analytic continuation.
 The spin chain sigma model actions will be written
in a manifestly $G$-invariant form in terms of the Cartan forms
and the string coordinates on
the supercoset $G/H$.

 The paper is organized as follows. In
Section \ref{scoh} we build the SU$(1,1|1)$ coherent state. In
Section \ref{sham} we evaluate the spin chain Hamiltonian in the
coherent state basis. By passing to the continuous limit and
expanding in derivatives we find a linear sigma model on the
group manifold $G/H$. In Subsection \ref{scartan} we rewrite the action in a
manifest SU$(1,1|1)$ form in terms of Cartan forms and the string
coordinates in the supercoset. In Section \ref{sstring} we show
that the same sigma model arises by considering a superstring
spinning fast on $S^1_{\phi}\times S^1_{\varphi}$ on AdS$_5\times S^5$. Finally in
Section \ref{scon} we summarize our results. Appendices \ref{aa},\ref{spin}
collect technical details and useful formulas.

\section{The coherent state}
\label{scoh}

In this section we derive a coherent state representation for the
 spin chain Hamiltonian with symmetry group SU$(1,1|1)$.
Coherent states are defined by the choice of a group $G$
and a state  $\ket{S}$ in a representation ${\cal R}$ of
the group. We denote by $H$ the stabilizer subgroup, \emph{i.e.} the group
of elements of $G$ that leave invariant $\ket{S}$ up to
a phase. The coherent state is then defined by
 the action of a finite group element of $g\in G/H$  on $\ket{S}$.

 We will take $\ket{S}$ to be the physical vacuum $\ket{\phi_0}$ and denote
by $G$ the rank two supergroup SU$(1,1|1)$.
  The generators for the algebra $\alG$ are taken to be
$$
T_A=(P_0,J_0,P,K,Q,S,\bar{Q},\bar{S} )~.
$$
 Conventions and details about the algebra and its singleton representation
 are given in appendix A.
 The stabilizer subgroup $H$ is generated by
$$
H=\{ P_0,J_0,Q,S \}={\rm SU}(1|1)\times {\rm U}(1)
$$
The coherent state is defined by acting with an element $g \in
G/H$ on the physical vacuum \bea\label{coh-st:gen}
\ket{\vec{n}}&=&g(\vec{n})\ket{\phi_0}=\e^{z P- \bar{z} K}\,
\e^{-\xi \bar{Q}-\bar{\xi} \bar{S}}\, \ket{\phi_0} \qquad\qquad
  \quad ~~~z=\rho\, \e^{2 \ii\,\phi}~
\eea
and it is parameterized by two real
parameters $\rho$ and $\phi$ and one complex grassmanian $\xi$.

Using \eqref{action:gen}, coherent states can be expanded in
 the basis $\{\ket{\phi_m},\ket{\lambda_m}\}$ , with
 $\phi_0$ a scalar field, $\lambda_0$ a fermion, and
 $m=0,1,2,\ldots $ labelling the number of derivatives.
  The expansion coefficients are given by~
\begin{equation}\label{cm}
  \ket{\vec{n}}=\sum_{m=0}^{\infty}\frac{\e^{2\ii\,\,m\,\phi}\,
\tanh^m\!\rho}{\cosh\!\rho}
  \left[ (1+\ft12\,\xi\bar{\xi})\,\ket{\phi_m}-{\xi\over \cosh\!\rho}
\,\ket{\lambda_m}\right]~.
\end{equation}
The expansion is such that
$$\bra{\vec{n}\, }\vec{n}\rangle=1~,$$
%%%%%%%%%%%%%%%%%%%%%%%%%%%%%%%%%%%%%%%%%%%%%%%%%%%%%%%%%%%%%%%%%%%%
%%%%%%%%%%%%%%%%%%%%%%%%%%%%%%%%%%%%%%%%%%%%%%%%%%%%%%%%%%%%%%%%%%%%
and the coherent states are over-complete
\begin{equation}\label{res-un}
  \I=\frac{2j+F}{\pi}\int_0^\pi\dd\phi\int_0^\infty\sinh\!2\rho\,\dd\rho\int\dd\xi\dd\bar{\xi}\,
  \ket{\vec{n}}\bra{\vec{n}}
~,
\end{equation}
%%%%%%%%%%%%%%%%%%%%%%%%%%%%%%%%%%%%%%%%%%%%%%%%%%%%%%%%%%%%%%%%%%%%
%%%%%%%%%%%%%%%%%%%%%%%%%%%%%%%%%%%%%%%%%%%%%%%%%%%%%%%%%%%%%%%%%%%%
with $F=0,1$ the supersymmetric grading of the state on which $\I$
acts. In the singleton representation   $j=0$ ,
formula (\ref{res-un})
%on bosonic states $|\phi_k\rangle$
is defined only in
the limit $j\rightarrow0$ (see appendix \ref{aa} and \cite{Bellucci:2004qr}
for details).

Conversely, to each coherent state $\ket{\vec{n}}$ we can
associate a point $n_A$ in the superspace
\be
 n_A
\equiv\bra{\vec{n} }\, T_A \, \ket{\vec{n} }~.\label{na}
\ee
Evaluating (\ref{na}) (charges $T_A$ are displayed in (\ref{action:gen}))
 one finds \bea
\left\{\begin{array}{rl}
n_{P_0}=&\ft12\,(1- \xi\bar{\xi})\, \cosh\!2\rho \\[3mm]
n_{J_0}=&\ft12\,(1+ \xi\bar{\xi})   \\[3mm]
n_{P}=& \overline{n_K}=~\ft{1}{2}\,\e^{-\ii\,2\phi}\,
\left(1 -\xi\bar{\xi}\right) \, \sinh\!2\rho\\[3mm]
n_{Q}=& \overline{n_S}=~  \e^{-\ii\,2\phi}\,\xi \,
\sinh\!\rho\\[3mm]
n_{\bar{Q}}=& \overline{n_{\bar{S}}}= \bar{\xi}\, \cosh\! \rho
\end{array}\right.\label{ncomp}
\eea
  The resulting vector is null $n^A n_A=0$ with respect to the
  Killing metric $g_{AB}$ defined by\footnote{
More precisely $ g_{AB}= \sum_{C,D}\, (-1)^{F_D}\, f_{A C}{}^D\, f_{B D}{}^C$
with $F_C=0,1$ depending whether the generator $C$ is even or odd
with respect to the supersymmetric grading. The inverse of the Killing metric also defines the Casimir as
$\hat{C}_2\equiv g^{AB}\, T_A T_B$ (see (\ref{casimir})).
}
\be
n_A m^A=g^{AB} n_A m_B= \ft12\,n_{P_0}\,m_{P_0}-
\ft12\, n_{J_0}\,m_{J_0}
-\ft12\,n_{P}\,m_{K}-\ft12\,n_{Q}\,m_S-\ft12\,\,n_{\bar{Q}}\,m_{\bar{S}}
+{\rm h.c.}
\label{metab}
\ee
 with $g^{AB}$ denoting the inverse of $g_{AB}$.

\section{Hamiltonian in the coherent state basis}
\label{sham}

Here we compute the average of the spin chain
 Hamiltonian over a coherent spin chain configuration
\be
 \ket{{\bf n}}=\ket{\vec{n}_1}\otimes \ket{\vec{n}_2}\ldots \ket{\vec{n}_J}
\label{kcoh}
\ee
 with $\ket{\vec{n}_k}$ denoting the coherent state describing the spin chain
excitation at site $k$ and time $t$.

 The spin chain action is given in terms of the spin chain Hamiltonian $H$
 by \cite{Bellucci:2004qr}
\bea
S &=& -\int \dd t \left( \ii\,\bra{\bf n} \partial_t \ket{\bf n}+
{\hat \lambda} \bra{\bf n} H \ket{\bf n}   \right).\label{action}
\eea
 The first (Wess-Zumino) term can be easily evaluated by taking the derivative of (\ref{cm})
 and then performing
the infinite sum. It has the simple form
\bea
\ii \bra{\bf n} \partial_t\! \ket{\bf n} &=&
\sum_k \left[-C_t+\ft{\ii}{2}(\bar{\xi}\,D_t\xi+\xi\,\bar{D}_t\bar{\xi})
\right]_k\label{ndn}\\
D_t &\equiv& \pd_t+\ii\, C_t~,
\quad\quad C_t\equiv 2\, \sinh^2\!{\rho}\ \partial_t
\phi~. \label{WSZ0}
\eea
  Evaluating the second term requires more work.
  The first task is to rewrite the SU$(1,1|1)$ two-site harmonic Hamiltonian
\cite{Beisert:2003jj} in our $(\phi_m,\lambda_m)$ basis.
 One finds
$$
  H=\sum_{k=1}^{J}H_{k\,k+1}
$$
with
\begin{equation}
\begin{array}{lll}
H_{12}\ket{A_k,B_l} &=&\dst
   \Big(h(k+\alpha)+h(l+\beta)\Big)\ket{A_k,B_l}-\left(\ft{\beta(1-\alpha)}{k+1}+\ft{\alpha(1-\beta)}{l+1}\right)\ket{B_k,A_l}\\
&&           \dst -\sum_{i=1}^k\left(\ft1i-\ft{\beta}{l+i+1}\right)\ket{A_{k-i},B_{l+i}}-\sum_{i=1}^k\ft{\alpha(1-\beta)}{l+i+1}\ket{B_{k-i},A_{l+i}}\\[5mm]
&&           \dst -\sum_{i=1}^l\left(\ft1i-\ft{\alpha}{k+i+1}\right)\ket{A_{k+i},B_{l-i}}-\sum_{i=1}^l\ft{\beta(1-\alpha)}{k+i+1}\ket{B_{k+i},A_{l-i}}~,\\[5mm]
  \label{Ham:pair}
\end{array}\end{equation}
and
$$ \ket{A,B}\equiv\ket{A}\otimes\ket{B}~~,
\qquad \alpha\equiv\delta_{A=\lambda},~~\beta\equiv\delta_{B=\lambda}~~,
\qquad h(m)=\sum_{i=1}^m\frac1i~~~~.
$$
The Hamiltonian (\ref{Ham:pair}) has a nice representation in the
coherent state basis $\ket{\vec{n}}$. We first compute the average of $H_{k\, k+1}$ over two-site
coherent states $\ket{\vec{n}_k \vec{n}_{k+1}} \equiv
\ket{\vec{n}_k}\otimes \ket{\vec{n}_{k+1}}$, then we sum up over
the spin chain sites $k=1,\ldots J$. The algebra is extremely long but the
result can be written in the remarkably simple form~
\bea\label{Ham:final}
   \bra{\bf n} H \ket{\bf n} &=&\sum_{k=1}^J\,  \bra{\vec{n}_k \vec{n}_{k+1}} \, H_{k\,k+1}\, \ket{\vec{n}_k \vec{n}_{k+1}}
  = \sum_{k=1}^J\, \log\left[1+2\,\vec{n}_k.\vec{n}_{k+1}\right]\nn\\
 & =&
\sum_{k=1}^J\, \log\left[1-\left(\vec{n}_k-\vec{n}_{k+1}\right)^2\right]~.
\eea
 As before, we use the Killing metric (\ref{metab}) to compute
  the scalar products in (\ref{Ham:final})\footnote{Notice that
  the scalar product here is not positively defined since the group
  is non-compact.}.

The coherent state representation (\ref{Ham:final}) is the
main result of this section. Note that in terms of
$\vec{n}_k$, it takes exactly the same
form as for the $sl(2)$ case, but now in terms of
the $SU(1,1|1)$ vector and the corresponding
Killing metric.

In the continuous limit, (\ref{Ham:final}) reduces to
\be
\begin{array}{ll}
\bra{\bf n} H \ket{\bf n} &= -\dst{1\over J} \int \dd\sigma\, \dst\,
 g^{AB}\, \partial_\sigma n_A \, \partial_\sigma n_B \\[5mm]
 &=\dst{1\over J} \int \dd\sigma\,\Bigl(\bar{D}_\sigma\bar{\xi}D_\sigma\xi+\left(1+\bar{\xi}\xi\right)\left[
(\partial_\sigma \rho )^{2} + \sinh^2 \!2\rho\ (\partial_\sigma\phi
)^{2}\right]\Bigr)
 \end{array}\label{h2}
\ee
 and coincides with the Hamiltonian of a string moving on the supercoset
 manifold $G/H$.

Finally, plugging (\ref{ndn}) and (\ref{h2}) into (\ref{action}),
one finds for the spin chain action the result
\bea
 S & =& -J\,\int \dd\sigma\dd t \left(-C_t+
 \ii \,\overline{\xi } D_t \xi
+{\hat{\lambda}\over J^2} \, ( {\bf e}^2
+ \bar{D}_\sigma\overline{\xi} \, D_\sigma \xi +
\overline{\xi }\,\xi\, {\bf e}^2)\right),\label{finalspinS}
\eea
with
\bea
{\bf e}^2 &=&(\partial_\sigma \rho )^{2} +\sinh^2\!2
\rho\ (\partial_\sigma\phi )^{2}~.\nn\\
D_a &=&\partial_a+\ii\,C_a \quad\quad C_a\equiv
2\,\sinh^2\!\rho\,\partial_a \phi~.\label{covD}
\eea
In Section \ref{sstring} we will find that the same sigma model
describe strings spinning fast on $S_\phi^1\times S_\varphi^1$ inside $AdS_5\times S^5$.

\subsection{Cartan forms}
\label{scartan}

The result (\ref{finalspinS}) can be written in a manifestly $G$-invariant
form in terms of the Cartan forms $L^A$ and the string coordinates $n_A$
in the supercoset $G/H$. This is the aim of this subsection. Readers not
interested in these details can skip this section.

 The Cartan forms $L^A$ are defined by
\be
\dd g\, g^{-1}=L^A\, T_A=L_a^A\, T_A\, \dd \sigma^a\ ,\quad\quad
\sigma^{0,1}\equiv(t,\sigma)~,
\label{defg}
\ee
with $g$ given by (\ref{coh-st:gen}).
They parameterize the gradient $\dd n_A =\partial_a n_A\, \dd \sigma^a$
of the string position on the supercoset along its worldsheet
coordinates
 $\sigma^a$. The explicit relation can be determined
as follows:
\bea
\dd n_A &=&\langle 0 |\,\dd g^{-1}\, T_A \,g\, |0\rangle+
\langle 0 |\, g^{-1}\, T_A \,\dd g\, |0\rangle \nn\\
&=&-L^B \, \langle \vec{n}|\,\{ T_B, T_A ]  \, |\vec{n}\rangle \nn\\
 &=&- L^B\,f_{BA}}^C\,  {n_C~, \label {ln}
\eea
 with $\{ ]$ denoting commutators or anticommutators according to
 spin of the generator.
 The first term in (\ref{action}) can also be written in
terms of $L^A$ and $n_A$
\be
\langle \vec{n}|\, \partial_t |\vec{n}\rangle = \langle \vec{n}|\,
\partial_t g\, g^{-1} |\vec{n} \rangle=L_t^A \,n_A. \label{swz}
\ee
 Plugging (\ref{ln},\ref{swz}) in (\ref{action}) one can finally rewrite
the spin chain action in a manifestly G-invariant form
\be
 S= -J\,\int \dd ^2\sigma\left[\ii\,L^A_t\, n_A-
 {\hat{\lambda}\over J^2}\, (L_{\sigma}^B\, {f_{BA}}^C\, n_C )^2\right].\label{act}
\ee
 This formula is the main result of this section and it is valid for
 any spin chain with Hamiltonian given by the first line in
(\ref{h2}) !

 In the present case one finds
\be\left\{
\begin{array}{ll}
L^{Q} &= \overline{L^{S}}={\e}^{2\ii \,\phi }\,
\sinh\! \rho\ \dd\overline{\xi } \\[3mm]
 L^{\bar{Q}}&= \overline{L^{\bar{S}}}=-\cosh\!\rho\  \dd\xi \\[3mm]
L^{P} &= -\overline{L^{K}}= {\e}^{2\ii \,\phi }
\, \left [\dd\rho+\sinh\!2\rho\,\left(\ii \,\dd\phi-
\ft14 \,\dd\overline{\xi }\, \xi+\ft14 \,\overline{\xi }\,
\dd\xi\right)\right]\\[3mm]
L^{P_0}&= -4\,\ii\, \sinh^2\!\rho\ \dd\phi-\ft12\cosh\!2\rho\,
(\overline{\xi }\, \dd\xi- \dd\overline{\xi }\, \xi)\\[3mm]
L^{J_0}&= \ft12\, (\dd\overline{\xi }\, \xi-\overline{\xi }\,
\dd\xi)
\end{array}\right.~.\label{lcomp}\ee
and the result (\ref{finalspinS}) follows.

\section{String action}
\label{sstring}

Here we describe the string duals of the
SU$(1,1|1)$ spin chain system. We follow the strategy
sketched in
\cite{Hernandez:2004kr} for superstrings spinning on a
sphere. The results will be related to that case
 via analytic continuation to AdS.

 To second order in the fermionic excitations, the string action on
AdS$_5\times S^5$ is given  by \cite{Metsaev:1998it} \be
 S=
 {R^2\over 4 \pi \alpha^\prime}
 \,\int\, \dd \sigma \dd\tau \,
\left[ g_{MN}\, \partial^a X^M\,
\partial_a X^N-2\, \ii\, \bar{\vartheta}\,
(\rho^a {\cal D}_a+\ft{\ii}{2}\, \epsilon^{ab}\,
\rho_a\, \Gamma_*\, \rho_b)\,\vartheta
\right]
\ee
with
\be
\rho_a=\partial_a X^M \, E_M^A\, \Gamma_A\,, \quad \quad
{\cal D}_a\equiv\partial_a+A_a\,, \quad\quad
A_a\equiv\ft14 \partial_a X^M \, \omega_M^{AB}\, \Gamma_{AB}\,.\label{defs}
\ee
Here
$E_M^A$ is the Zehnbein, $\Gamma^A$ are the usual flat
ten-dimensional gamma matrices, $\Gamma_*$ is the chirality operator
along AdS, $\rho_a$ is the induced gamma matrices and $\omega_M^{AB}$ is
the spin connection.
We first write the metric for the SL$(2)$ spinning string as
\begin{equation}
  \dd s^2=-\cosh\!^2\rho\ \dd \hat{t}^2+{\dd \rho}^{2}+\dd \hat{\varphi}^2+\sinh^2\!
  \rho\  {\dd \hat{\phi}}^{2} \label{metr0}
\end{equation}
with $\hat{t},\rho,\hat{\phi}$ denoting three coordinates inside AdS$_5$ and
$\varphi_3$ being an angle on $S^5$.
 Tangent space labels $A=0,1,2,3$ will be associated with the
coordinates $\hat{t},\rho,\hat{\varphi},\hat{\phi}$ respectively.
 In terms of
such variables, one has $\Gamma_*\equiv\ii\,\Gamma_{01345}$.
 We introduce the notation $\bar{\Pi}\equiv\Gamma_{45}$ and choose our
 ten-dimensional spinors
 such that $\bar{\Pi}=\ii $.

 Then, we make the change of
coordinates
$$\hat{t}\rightarrow t-\varphi~,\quad\hat{\varphi}\rightarrow t~, \quad \hat{\phi}\rightarrow t-\varphi+2\,\phi$$
in order to bring the metric into a form with $g_{tt}=0$ where a BMN like
limit (see below) is well defined\footnote{We will
not need the explicit form of this metric here. Spin connections will be
computed using the starting metric (\ref{metr0}).}.
We take
\be
t=\kappa\,\tau~,\quad\quad\dot{X}^{M}\equiv\pd_t X^M\quad\quad
X^{M'}\equiv\pd_\sigma X^M
\ee
and consider the limit $\kappa\to \infty$
keeping $\kappa^2\,\dot{X}^{M\neq t}$ fixed \cite{Kruczenski:2003gt}.

We look for superstring excitations satisfying
the Virasoro constraints \begin{eqnarray} &&g_{MN} \,\partial_\tau
X^M
\partial_\sigma
 X^N-\ii \,
\bar{\vartheta}\,(\rho_\sigma {\cal D}_\tau+\rho_\tau
{\cal D}_\sigma)\vartheta=0\nn\\
&&g_{MN}\, (\partial_\tau X^M  \partial_\tau
 X^N+\partial_\sigma X^M \partial_\sigma X^N)
 -2\, i\,\bar{\vartheta}\,(\rho_\tau {\cal D}_\tau
 +\rho_\sigma {\cal D}_\sigma)\vartheta=0.\nn
\end{eqnarray}

They can be used to solve for $\varphi$ in favor of the remaining
variables. To leading order in $\kappa$ one finds
\bea
\varphi' &=&-2\,\sinh^2\!\rho\ \phi'+{\cal O}(\vartheta^2)\nn\\
\dot{\varphi} &=&-2\,\sinh^2\!\rho\ \dot{\phi}-{{\bf e}^2\over
{2\,\kappa^2}}+{\cal O}(\vartheta^2)\label{vir}
\eea
with ${\bf e}^2 =\rho'^{2} +\sinh^2\!2
\rho \ \phi^{'2}$.
One can easily see that fermionic terms in (\ref{vir}) contribute to the lagrangian
 either as quartic terms in the fermions or to
 subleading terms in the $1\over\kappa$-expansion and therefore can be
 discarded.

Using the first of these equations we can write the bosonic part of the action as
\be
S_B=-{R^2 \kappa\over 2 \pi \alpha'}\,\int \, \dd\sigma\dd t \,\left(-\dot{\varphi}- C_t+{{\bf e}^2\over
{2\,\kappa^2}}\right)\label{finalstringSB}
\ee

Now let us consider the fermionic Lagrangian.
 Evaluating (\ref{defs}) one finds
\bea
\rho_0&=&-\kappa\,\left [\left ( 1-\dot{\varphi}\right )\,\cosh\!\rho \, \Gamma_0
- \dot{\rho} \,         \Gamma_1
 - \Gamma_2
 -\left ( 1+2\,\dot{\phi}-\dot{\varphi}\right )  \, \sinh\!\rho \, \Gamma_3 \right ] \label{rho0}\\
\rho_1&=&\varphi' \, \cosh\!\rho\ \Gamma_0 +\rho' \, \Gamma_1
+\left ( 2\,\phi'-\varphi'\right )  \, \sinh\!\rho\ \Gamma_3  \nn\\
A_0&=& \ft\kappa2\left[\left (- 1+\dot{\varphi}\right )\,
\sinh\!\rho \ \Gamma_{01}
 -\left (1+2\,\dot{\phi}-\dot{\varphi}\right )
 \, \cosh\!\rho\ \Gamma_{13}\right]\nn\\
A_1&=&\ft12\,\varphi' \, \sinh\!\rho\ \Gamma_{01}
-\left (\phi'-\ft12\,\varphi'\right )  \, \cosh\!\rho\
\Gamma_{13}\nn
\eea

 In addition, eqs. (\ref{vir}) can be used to show that the matrices $\rho_a$ satisfy the
Clifford algebra \bea -\{ \rho_0,\rho_0 \}&=& \{ \rho_1,\rho_1 \}=
2\,\partial_{\sigma} X^M\, \partial_{\sigma} X_M=2\,{\bf e}^2  \nn\\
\{ \rho_1,\rho_0 \}&=&
2\,\partial_{\sigma} X^M\, \partial_{\tau} X_M=0
\eea
 and therefore can be put in the form\footnote{As it can be seen
 from the coefficient in front of $\Gamma_0$ in eq. (\ref{rho0}), $\rho_0$ gets a negative sign.}
\be
\rho_0=-{\bf e}\ \Gamma_0\,, \quad \rho_{1}={\bf e}\ \Gamma_3\,,
\label{actf}
\ee
via a spinor rotation $\rho_a\to S \, \rho_a \, S^{-1}$.
The precise form of $S$ and its derivation are given in Appendix \ref{spin}.
 In the new basis the fermionic string action reads
\bea
 S_F=  \ii\,{R^2 \kappa\over 4 \pi \alpha'}\,\int\, \dd\sigma\dd t \,\bar{\Psi}\, \left(\Gamma_0 {\cal D}_t
+\frac1\kappa\,\Gamma_3 {\cal D}_\sigma +\left(1+{{\bf e}^2\over
{2\,k^2}}\right)\ii\, \Gamma_{1} \right)\Psi \label{actt} \eea
where $${\cal D}_a=\pd_a+C_a\,\Gamma_{0123}~,$$ with $C_a$ defined
as in (\ref{covD}). In the derivation of (\ref{actt}) the
field $\phi$ is taken to be on the mass shell up to
order\footnote{We believe that this is also the case for
the $SU(1|2)$ spin chain considered in \cite{Hernandez:2004kr}.} ${1\over k^2}$.
This is consistent with the fact that
the limit $\kappa\to \infty$ in (\ref{actt}) corresponds to the
semiclassical expansion around $\hbar\sim {1\over \kappa}\to 0$ where
fields are put on the mass shell.

Following \cite{Hernandez:2004kr} we choose our spinor as a four
dimensional Majorana spinor:
$$\Psi= \left(%
\begin{array}{c}
 e^{-i t} \,  \xi \\
 e^{i t} \,  \bar{\xi} \\
\end{array}%
\right)~\,
\quad
\bar{\Psi}=\Psi^\dagger \, \Gamma_0
\quad ~~~~~~\quad
\xi=\left(%
\begin{array}{c}
  \xi_1 \\
  \xi_2 \\
\end{array}%
\right)
\quad
\bar{\xi}=\left(%
\begin{array}{c}
  i\,\xi^*_2 \\
  -i\,\xi^*_1 \\
\end{array}%
\right)~.
$$
The factor $e^{i t}$ is included, in order to remove the
fast mode fermionic oscillations. For the Gamma matrices we take:
$$\Gamma_0= \left(%
\begin{array}{cc}
0 & \sigma_1 \\
-\sigma_1 & 0 \\
\end{array}%
\right)~\,
\quad
\Gamma_1= \left(%
\begin{array}{cc}
0 & \ii\,\sigma_2 \\
-\ii\, \sigma_2 & 0 \\
\end{array}%
\right)~\,
\quad
\Gamma_2= \left(%
\begin{array}{cc}
0 & -\ii\,\sigma_3 \\
\ii\,\sigma_3  & 0 \\
\end{array}%
\right)~\,
\quad
\Gamma_3= \left(%
\begin{array}{cc}
0 & -1 \\
-1 & 0 \\
\end{array}%
\right)~.
$$

Plugging into (\ref{actt})  one finds
the fermionic action: :
\bea
S_F &=& - {R^2 \kappa\over 2 \pi \alpha'}\,\int\, \dd\sigma\dd t
\,\left[\ii\, \xi_1^*
(D_t \xi_1+\frac1\kappa D_\sigma \xi_2)+\ii\,\xi_2^*
(D_t  \xi_2+\frac1\kappa D_\sigma \xi_1)
\right.
\nn\\
&&\left. ~~~~~~~~~~~~~~~~~~~
~~~~+{{\bf e}^2\over{2\,\kappa^2}}\,\xi^*_1 \xi_1-
(2+{{\bf e}^2\over{2\,\kappa^2}})\, \xi_2^*\xi_2\right]
\eea
with covariant derivatives given by (\ref{covD}).
 To leading order in ${1\over k}$ the field $\xi_2$ is non-dynamical and
 can be solved via its equation of motion in favor $\xi_1$.
$$
\xi_2={\ii\over 2 \kappa}\,D_\sigma\, \xi_1.
$$
Plugging into the action one finally finds
\bea
S_F &=& -{R^2 \kappa\over 2 \pi \alpha'}\,\int\, \dd\sigma\dd t  \,\left[
\ii\,\xi^*_1 D_t \xi_1
+\frac1{2\,\kappa^2}\left(D_\sigma \xi^*_1\,D_\sigma \xi_1+{\bf e}^2\,
\xi^*_1 \xi_1)\right)\right].\label{finalstringSF}
\eea
 One can easily see that the string
action $S=S_B+S_F$ following from
 (\ref{finalstringSB})+(\ref{finalstringSF})
perfectly matches the spin chain result
(\ref{finalspinS}) after the identifications \cite{Bellucci:2004qr}
\be
J={R^2 \kappa\over 2 \pi \alpha'}\quad\quad~~~~~~~~~
\hat{\lambda}={R^4 \over 8 \pi^2 \alpha'^2}
\ee

\section{Summary of results}
\label{scon}

 In this note we derive a coherent state representation for the integrable
 spin chain Hamiltonian with symmetry group $G=$SU$(1,1|1)$.
 The result can be cast in the remarkably compact and
 simple form
 $$
\bra{\bf n} H \ket{\bf n} =
\sum_{k=1}^L\, \log\left[1-\left(\vec{n}_{k+1}-\vec{n}_{k}\right)^2\right]
$$
 with $\vec{n}_k$ parameterizing a point in the supercoset
 $G/H$, $H=$SU$(1|1)\times $U$(1)$ being the stabilizer group.
  The scalar product is defined in terms of the Killing metric on $G$.

By passing to the continuous limit
$\partial_a \vec{n}\equiv(\vec{n}_{k+1}-\vec{n}_{k})J$, we find a spin chain sigma
model describing a string moving on the group manifold $G/H$ .
The result can be written in a manifestly $G$-invariant form in  terms of the
Cartan forms $L^A$ and the string coordinates $n_A=\bra{\bf n} T_A \ket{\bf n}$
on the supercoset
\bea
 S &=& -\int \dd^2 \sigma\, \left[\ii\,\bra{\bf n} \partial_t \ket{\bf n}
 +\hat{\lambda} \bra{\bf n} H \ket{\bf n}\right]=
- \int \dd^2 \sigma \left[\ii\,L^A_\tau\, n_A-
 {\hat{\lambda}\over J^2}\,
 %(L_{\sigma}^B\, f_{BA}{}^C\, n_C )^2
g^{AB} \partial_\sigma n_A\,\partial_\sigma n_A
 \right]~.\nn
 \eea
 Here $n_A(\sigma,t)$ describes the profile of the string evolving in time and $g_{AB}$ denotes
  the Killing metric of $SU(1,1|1)$. The same formula applies to $SU(2|3)$ \footnote{This can be
  easily seen for the $SU(1|2)$ case by replacing hyperbolic functions in (\ref{ncomp},\ref{lcomp}) by their trigonometric analogs  and
   comparing the resulting action with \cite{Hernandez:2004kr}}  .
  In components one finds
 \bea
S  & =& -J\,\int \dd\sigma\dd t \left[-C_t+\ii \,\overline{\xi }
D_t \xi
+{\hat{\lambda}\over J^2} \, ( {\bf e}^2
+ \bar{D}_\sigma\overline{\xi} \, D_\sigma \xi +
\overline{\xi }\,\xi\, {\bf e}^2)\right]~.
\nn
\eea
  The same sigma model was found by considering the Green-Schwarz (GS)
  action of a superstring
  spinning fast on a $S_\phi^1\times S_\varphi^1$ torus inside
  AdS$_5\times S^5$. This establishes a precise
  map between coherent states in the SU$(1,1|1)$ sector and string states
  and matches their dynamics.

  It is worth to stress that, unlike
  the GS action, the spin sigma model Lagrangian
  is built out of the $SU(1,1|1)$
  invariant Killing metric of the supergroup.
  The two actions are related in a highly non-trivial way in
  the limit $J\to \infty$ where the string becomes semiclassical
  and fields come near the mass shell. The analysis here
  provide us with a detailed dictionary between the two descriptions.
  In particular the agreement found here between the two actions
  implies a similar match between their classical solutions.
  It would be nice to explore the simpler spin chain sigma model
description  as a possible bottom-up definition for the study
of more general string configurations on $AdS_5\times S^5$.

\subsection*{Acknowledgements}
We thank R. Hernandez, A. Marrani and C. Sochichiu for discussions. This work
was partially supported by INTAS-00-00254 grant, INTAS-00-00262, RF
Presidential grants MD-252.2003.02, NS-1252.2003.2, INTAS grant
03-51-6346, RFBR-DFG grant 436 RYS 113/669/0-2, RFBR grant
03-02-16193 and the European Community's "Marie Curie Research Training
Network under contract MRTN-CT-2004-005104 Forces Universe".

\appendix
\section*{Appendix}
\renewcommand{\theequation}{A.0}

In this appendix we collect the commutation relations and details
on the ``singleton" representations of  the superalgebra
$\alG=\alSU(1,1|1)$. A singleton corresponds
to a subsector of the ${\cal N}=4$ SYM multiplet that closes under
$\alG$. Here we adopt the oscillator description (see \cite{Beisert:2003jj} for
details). In this formalism, elementary SYM fields (the singleton
of $\alPSU(2,2|4)$) are represented by acting on a Fock vacuum
$|0\rangle$ with bosonic $(a_\alpha, b_{\dot{\alpha}})$ and
fermionic oscillators $c_A$,
($\alpha,\dot{\alpha}=1,2, A=1,\ldots 4$). Physical states
 satisfy the condition
\begin{equation}
C=n_a-n_b+n_c =2 \label{c0}
\end{equation}
with $n_a,n_b,n_c$ denoting the number of oscillators of a given type.

 The closed subalgebras of $\alSU(2,2|4)$ are defined by
restricting the range of $\alpha,\dot{\alpha},A$.

\renewcommand{\theequation}{\Alph{section}.\arabic{equation}}
\section{$\alSU(1,1|1)$ algebra}
\label{aa}

 The algebra $su(1,1|1)$ is built in terms of bilinears  of two
bosonic ($a$, $b$) and one fermionic ($c_3$) oscillators.
 The physical vacuum can be taken to be  $|\phi_0
%%%%%%%%%%%%%%%%%%%%%%%%%%%%%%%%%%%%%%%%%%%%%%%%%%%%%%%%%%%%%%%%%%%%
%%%%%%%%%%%%%%%%%%%%%%%%%%%%%%%%%%%%%%%%%%%%%%%%%%%%%%%%%%%%%%%%%%%%
 \rangle
%%%%%%%%%%%%%%%%%%%%%%%%%%%%%%%%%%%%%%%%%%%%%%%%%%%%%%%%%%%%%%%%%%%%
%%%%%%%%%%%%%%%%%%%%%%%%%%%%%%%%%%%%%%%%%%%%%%%%%%%%%%%%%%%%%%%%%%%%
 =c^\dagger_1c^\dagger_2|0\rangle$.
 States (``letters") in the singleton representation are given by
\bea
 |\phi_m\rangle &=&\frac{1}{m!}(a\dg b\dg)^m |\phi_0\rangle
~~~~~~~~\Leftrightarrow {1\over m!}\, {\cal D}^m \phi \nn\\
 |\lambda_m\rangle &=&\frac{1}{m!}(a\dg b\dg)^m\, b\dg c_3\dg
|\phi_0\rangle ~~\Leftrightarrow {1\over m!}\,
 {\cal D}^m \lambda\label{stasu111}
\eea
 and correspond to a scalar field $\phi_0$, a fermion $\lambda_0$
and their $m$-derivatives along a fixed direction.
 The algebra in this case is
non-compact and the reprentations are infinite-dimensional.

The generators can be written as bilinears in the oscillators
 \bea \left.\begin{array}{l}
P=a\dg b\dg \\
K=a b\\
P_0=\ft12(1+a\dg a+b\dg b)\\
J_0=\ft12(1+a\dg a-b\dg b)
\end{array}
\right. \quad\quad
 \left. \begin{array}{l}
Q=a\dg c_3\\
\bar{Q}=b\dg c_3\dg\\
S=a c_3\dg\\
\bar{S}=b c_3
\end{array}\right. ~. \nn
\eea The charges $P_0$ and $J_0$ give the Cartan of the group. Non
vanishing commutation relations are given by
$$\begin{array}{lllllll}
[K,P]&=2P_0     \quad\quad\quad          &[S,P]&=\bar{Q}      \quad\quad\quad  & [K,\bar{Q}]&=S\\
\{Q,S\}&= P_0-J_0              &[\bar{S},P]&=Q                       & [K,Q]&=\bar{S}\\
\{\bar{Q},\bar{S}\}&= P_0+J_0   &\{\bar{Q},Q\}&=P
& \{\bar{S},S\}&=K\\
&                                 & [P_0,Q]&=\ft12\, Q      \quad\quad\quad  & [J_0,Q]&=\ft12\, Q \\

[P_0,K]&=-K                       & [P_0,\bar{Q}]&=\ft12\, \bar{Q}
        & [J_0,\bar{Q}]&=-\ft12\, \bar{Q} \\

[P_0,P]&=P     \quad\quad\quad  & [P_0,S]&=-\ft12\, S
     & [J_0,S]&=-\ft12 S \\

&                                 & [P_0,\bar{S}]&=-\ft12\,
\bar{S}
         & [J_0,\bar{S}]&=\ft12 \bar{S}
\end{array}~.$$
 The action of the generators on the states (\ref{stasu111}) is given by
\begin{align}\label{action:gen}
  P_0\ket{\phi_m}&=(m+\ft12)\ket{\phi_m}       &P_0\ket{\lambda_m}&=(m+1)\ket{\lambda_m}\nnb\\
  J_0\ket{\phi_m}&=\ft12\,\ket{\phi_m}              &J_0\ket{\lambda_m}&=0\nnb\\
  P\ket{\phi_m}&=(m+1)\ket{\phi_{m+1}}       &P\ket{\lambda_m}&=(m+1)\ket{\lambda_{m+1}}\nnb\\
  K\ket{\phi_m}&=m~\ket{\phi_{m-1}}          &K\ket{\lambda_m}&=(m+1)\ket{\lambda_{m-1}}\nnb\\
  Q\ket{\phi_m}&=0                           &Q\ket{\lambda_m}&=(m+1)\ket{\phi_{m+1}}\nnb\\
  S\ket{\phi_m}&=\ket{\lambda_{m-1}}         &S\ket{\lambda_m}&=0\nnb\\
  \bar{Q}\ket{\phi_m}&=\ket{\lambda_m}   &\bar{Q}\ket{\lambda_m}&=0\nnb\\
  \bar{S}\ket{\phi_m}&=0                     &\bar{S}\ket{\lambda_m}&=(m+1)\ket{\phi_m}.
\end{align}
For later convenience we choose the normalization
$$\left<\phi_m|\phi_n\right>=\delta_{mn}~,\quad \left<\lambda_m|\lambda_n\right>=(m+1) \delta_{mn}~,
\quad \left<\phi_m|\lambda_n\right>=0~.$$

 SYM operators in the $SU(1,1|1)$ sector are
given by tensor products of $J$ singletons ``words made out of letters'',
i.e. we take J copies of the considered oscillators $a,b,c$ and impose the condition
(\ref{c0}) at each site.
 The symmetry algebra is taken to be the diagonal SU$(1,1|1)$ algebra
\be
 T^A=\sum_{k=1}^J\, T^A_k
\ee
 with $\vec{T}_k^A$ acting on the $k^{th}$ site.

 It is not difficult to verify that the quadratic operator
\begin{equation}\label{casimir}
  \hat{C}_2= g^{AB}\,T_A T_B= P_0^2- J_0^2-\frac12\{P,K\}-\frac12[Q,S]-\frac12[\bar{Q},\bar{S}]
\end{equation}
commutes with all generators, i.e. it is a Casimir of the algebra.
Therefore, it is proportional to a unit matrix:
%%%%%%%%%%%%%%%%%%%%%%%%%%%%%%%%%%%%%%%%%%%%%%%%%%%%%%%%%%%%%%%%%%%%
%%%%%%%%%%%%%%%%%%%%%%%%%%%%%%%%%%%%%%%%%%%%%%%%%%%%%%%%%%%%%%%%%%%%
$$  \hat{C}_2=j~(j+1)~ \I~.$$
The number $j$ labels the irreducible representations of the algebra.
In particular, for $j=0$ the Casimir vanishes.
 The defining representation $j=0$ is the so called
``singleton representation'' and is generated by acting with the
lowering charges on the highest weight $|\phi_0\rangle \equiv
c_1\dg c_2\dg|0\rangle$.

All spin $j$ representations arise already in the tensor product
of two singletons.
 The spin $j$ highest weight state
spin $j$ representation can be written as follows: \be
 |j\rangle_{k_1 k_2}=\sum_{n=0}^j \,(-1)^n\,\left(\!\!
\begin{array}{c}
  j \\
  n \\
\end{array}%
\!\!\right)\, |\phi_{j-n}\rangle_{k_1} \otimes |\phi_{n}\rangle_{k_2}~.\label{spinj}
\ee
%%%%%%%%%%%%%%%%%%%%%%%%%%%%%%%%%%%%%%%%%%%%%%%%%%%%%%%%%%%%%%%%%%%%
%%%%%%%%%%%%%%%%%%%%%%%%%%%%%%%%%%%%%%%%%%%%%%%%%%%%%%%%%%%%%%%%%%%%

\section{Spinor rotations}
\label{spin}

  Here we derive the spinor rotation $S$ and string action
 in the new spinor basis. These results are relevant for the analysis of  Section \ref{sstring}.
  $S$ is defined by
  $$
  S \rho_0 S^{-1}=-{\bf e}\, \Gamma_0\quad \quad S \rho_1 S^{-1}={\bf e}\, \Gamma_3
  $$

   As in \cite{Hernandez:2004kr} we write $S$ as a product of rotations of the type
   \bea
   S_{ij}(p)&=& \e^{{1\over 2}p\ \Gamma_{ij}}=\cos{p\over 2}+\sin{p\over 2}\ \Gamma_{ij}\nn\\
    S_{0i}(p)&=& \e^{{1\over 2}p\ \Gamma_{0i}}=\cosh\!{p\over 2}+\sinh\!{p\over 2}\ \Gamma_{ij}.\nn
   \eea

 The spinor rotation can be written as
  \be
   S=S_{13}(p_4)\,S_{02}(p_3) \,S_{01}(p_2)\,S_{03}(p_1)
   \label{sp}
\ee
 with%\footnote{Notice that the $p_4$-rotation is defined only when
 %$-\sinh\!2\rho \ {\phi'\over {\bf e}}$ is bigger than one. We will assume here that this is always
 %the case.}

 \begin{align}
 p_1&=\rho+ \sinh\!2\rho \ \dot{\phi}  &p_2& =\dot{\rho} \nn\\
 \cosh p_3& = {\kappa\over {\bf e}}+{{\bf e}\over {2\,\kappa}}
 & \cos p_4&=\sinh\!2\rho \ {\phi'\over {\bf e}}\nn\\
 \sinh p_3& ={\kappa\over {\bf e}}
 &\sin p_4& =-{\rho'\over {\bf e}}.
 \end{align}
 The transformed matrices can be computed with the help of
\bea
S_{ij} \, \Gamma_i\, S_{ij}^{-1}&=&\cos p\, \Gamma_i-\sin p \, \Gamma_j\nn\\
S_{ij} \, \Gamma_j\, S_{ij}^{-1}&=&\sin p \, \Gamma_i+\cos p\, \Gamma_j\nn\\
S_{0i} \, \Gamma_0\, S_{0i}^{-1}&=&\cosh\! p\, \Gamma_0+\sinh\! p \, \Gamma_i\\
S_{0i} \, \Gamma_i\, S_{0i}^{-1}&=&\sinh\! p \, \Gamma_0+\cosh\! p\, \Gamma_i~.\nn
\eea
 In the process, we use the Virasoro constraints
\be
\varphi'+2\,\sinh^2\!\rho\ \phi' =0\ ,\quad\quad
\dot{\varphi}+2\,\sinh^2\!\rho\ \dot{\phi} =-{{\bf e}^2\over {2\,\kappa^2}}
\ee
in order to solve for $\varphi$ in  favor of $\phi$. We also use
the equality given by the cross derivative condition
$\pd_t\varphi'=\pd_\sigma\dot{\varphi}$ to solve for $\rho''$.

At intermediate steps, one gets
\bea
\ft\ii2\,\epsilon^{ab}\,S\,\rho_a \, \Gamma_* \,\rho_b\, S^{-1}&=&
-\kappa\,{\bf e}\,\left ( 1+\frac{{\bf e}^2}{2\,\kappa^2}\right)\,\Gamma_1\,\bar{\Pi}\nn\\[5mm]
S\,A_\tau\,S^{-1}&=&\frac{\kappa^{2}}{2 \, {\bf e}^{2}}\ \left(\dot{\rho}\,\rho'+ 2 \, \sinh^{2}\!2 \rho \  \dot{\phi}\, \phi'
\right)\left(\Gamma_{01}-\Gamma_{12}\right)\nn\\
&&-\frac{\kappa^{2} }{2 \, {\bf e}^{2}}\, \sinh\!2 \rho \left(\dot{\rho} \, \phi'-2 \, \dot{\phi} \, \rho'
\right)\left(\Gamma_{03}+\Gamma_{23}\right)\nn\\
&&-\frac\kappa2\,\left(1+\frac{{\bf e}^{2}}{2 \, \kappa^2}+2\, \cosh\!2\rho\,\dot{\phi}\right)\,\Gamma_{13}\nn\\[5mm]
S\,A_\sigma\,S^{-1}&=&\frac{\kappa }{{\bf e}^{2}}\left[  \cosh\!2\rho \,
\left (\dot{\rho} \, \rho'+\sinh^2\!2\rho\  \dot{\phi}
\,\phi'\right)\,\phi'+\ft12 \sinh^2\!2\rho\ \phi'^2\right]
 \left(\Gamma_{01}-\Gamma_{12}\right)\nn\\
&&+\frac{\kappa  }{{\bf e}^{2}} \, \sinh\!2\rho \
\phi' \left[ \ft12\rho'+\cosh\!2\rho  \,
\left (\dot{\phi} \, \rho'-\dot{\rho} \,\phi'\right ) \right]\left(\Gamma_{03}+\Gamma_{23}\right)\nn\\
&&-\cosh\!2\rho\ \phi'\,\Gamma_{13}
+\frac{1}{4 \, \kappa}\sinh^{2}\!2\rho\
\phi'^{2}\,\Gamma_{01}+\frac{1}{4\, \kappa}\, \sinh\! 2\rho  \ \rho' \,
\phi'\,\Gamma_{03}\nn\\
&&
\eea
and
\bea
S\,\pd_\tau\,S^{-1}&=&\frac{\kappa^{2}}{2 \, {\bf e}^{2}}\dot{\rho}  \, \rho' \left(\Gamma_{01}-\Gamma_{12}\right)
-\frac{\kappa^{2}}{2 \, {\bf e}^{2}}\, \sinh\!2\rho\ \dot{\rho}  \, \phi' \left(\Gamma_{03}+\Gamma_{23}\right)\nn\\
&&-\frac{\kappa}{2 \, {\bf e}^{2}}\left(2\cosh\!2\rho\ \dot{\rho} \, \rho' \, \phi'
+\left (\dot{\phi}'  \, \rho'-\dot{\rho}'  \, \phi'\right )  \, \sinh\!2\rho\right)\Gamma_{13}\nn\\[5mm]
%%%%%%%%%%%%%%%%%%%%%%%%%%%%%%%%%%%%%%%%%%%%%%%%%%%%%%%%%%%%%%%%%%%%
%%%%%%%%%%%%%%%%%%%%%%%%%%%%%%%%%%%%%%%%%%%%%%%%%%%%%%%%%%%%%%%%%%%%
&&+\frac{\kappa}{2 \, {\bf e}^{2}}\left(\dot{\rho}'\,\rho'+\sinh^2\!2\rho\ \dot{\phi}'\,\phi'
+\sinh\!4\rho\  \dot{\rho}\,\phi'^2\right)\Gamma_{02}\nn\\[5mm]
%%%%%%%%%%%%%%%%%%%%%%%%%%%%%%%%%%%%%%%%%%%%%%%%%%%%%%%%%%%%%%%%%%%%
%%%%%%%%%%%%%%%%%%%%%%%%%%%%%%%%%%%%%%%%%%%%%%%%%%%%%%%%%%%%%%%%%%%%
%
S\,\pd_\sigma\,S^{-1}&=&\frac{1}{4\, \kappa}\,\rho'\left(\rho'\,\Gamma_{01}-\sinh\!2\rho\
\phi'\,\Gamma_{03}\right)\nn\\
&&-\frac{\kappa^{2} }{{\bf e}^{2}}\, \left (\dot{\phi} \,\rho'-\dot{\rho} \,\phi'\right )  \, \sinh\!2\rho
\left(\Gamma_{02}+\sinh\!2\rho\ \frac{\phi'}{\rho'}\Gamma_{13}\right)\nn\\
&&- \left (\cosh\!2\rho\ \phi' +\ft12\sinh\!2\rho \,\frac{\phi''  }{\rho'}\right ) \,\Gamma_{13}\nn\\
&& +\frac{\kappa}{2\,{\bf e}^{2}}
\left(\rho'^{2}+2\,\cosh\!2\rho \ \dot{\phi} \,\rho'^{2}+\sinh\!2\rho \,\left (\dot{\phi}'  \,\rho'-\dot{\rho}' \,\phi'\right )
\right)\left(\Gamma_{01}-\Gamma_{12}\right)\nn\\
&&-\frac{\kappa}{2 \, {\bf e}^{2}}\left(\dot{\rho}'\,\rho'+ \sinh\!2\rho\ \phi' \, \left (\rho'+ 2\,\cosh\! 2\rho \ \dot{\phi}\,\rho'
+\sinh\!2 \rho \ \dot{\phi}' \right )
\right)\nn\\
&&\phantom{\frac{\kappa}{2 \, {\bf e}^{2}}}\times\left(\Gamma_{03}+\Gamma_{23}\right)
\eea

Rewriting the action  in terms of
$$\vartheta\equiv \sqrt{\frac{\kappa}{2\,\bf e}}\ S^{-1}\Psi~,\quad\quad\bar{\vartheta}=\sqrt{\frac{\kappa}{2\,\bf e}}\ \bar{\Psi}\,S~,$$
one finds
\begin{eqnarray}
S_F&=&\ii\,{R^2 \kappa\over 4 \pi \alpha'}\,\int\dd \sigma \dd\tau\,\bar{\Psi}\left[
\Gamma_0\left(\pd_\tau+S\,A_\tau\,S^{-1}+S\,\pd_\tau\, S^{-1}
+\sqrt{\bf e}\,\pd_\tau\frac1{\sqrt{\bf e}}\right)\right.\nn\\
&&\phantom{- 2\,\ii\,\kappa\,\int\dd^2 \sigma
}\left.+\Gamma_3\left(\pd_\sigma+S\,A_\sigma\,S^{-1}+S\,\pd_\sigma S^{-1}+\sqrt{\bf e}\,\pd_\sigma\frac1{\sqrt{\bf e}}\right)
+k\,\left ( 1+\frac{{\bf e}^2}{2\,\kappa^2}\right
)\,\Gamma_{1}\,\bar{\Pi}\right]\Psi\nn\\[8mm]
&=& \ii\,{R^2 \kappa\over 4 \pi \alpha'}\,\,\int\dd\sigma  \dd t \,\bar{\Psi}\left[\Gamma_0
\left(\pd_t+\left(C_t+\ft12\dot{p_4}+\dot{\phi}+\ft12\right)\Gamma_{0123}\right)\right.\nn\\
&&\phantom{- 2\,\ii\,\kappa^2\,\int}
 \left.+\frac1\kappa \Gamma_3\left(\pd_\sigma+\left(C_\sigma+
 \ft12 p_4'+\phi'+X\right)\Gamma_{0123}+X\,\Gamma_{13}\right)+\left ( 1+\frac{{\bf e}^2}{2\,\kappa^2}\right
)\,\ii\,\Gamma_{1}\right]\Psi\nn
\end{eqnarray}
with
$$X= \kappa^{2}\,\frac{\dot{\rho}}{\rho'}-2\,\cosh\!2\rho \ \phi'
-\sinh\!2\rho \ \frac{\phi''}{2 \, \rho'}=0+{\cal O}({1\over\kappa^2})~.$$
The left hand side here is proportional to the equation of motion for $\phi$
that should be satisfied to order ${1\over k^2}$ in the semiclassical limit $\hbar={1\over k}\to 0$.

In order to get rid of full derivatives in the connections and obtain (\ref{actt}), one can finally
 make the following change in the spinors:
$$\Psi\longrightarrow\e^{-\ft12(t+p_4+2\phi)\Gamma_{0123}}\Psi~.$$
% ---------------------------------------------------------------
%\bibliographystyle{JHEP-2}
%\bibliographystyle{amsplain}
%\bibliography{syst,syst2,ref}
% ---------------------------------------------------------------

\providecommand{\href}[2]{#2}\begingroup\raggedright\endgroup

\end{document}